\documentclass{iau} 
\usepackage{graphicx}

\title[IAUS~315.~~Gas flows in galactic nuclei] 
{Gas flows in galactic nuclei: observational constraints on BH-galaxy coevolution}

\author[S.~Garc\'{\i}a-Burillo]   
{Santiago Garc\'{\i}a-Burillo $^1$}

\affiliation{$^1$Observatorio de Madrid (OAN-IGN), \\ Alfonso XII, 3, 
Madrid-28014, Spain \\ email: {\tt s.gburillo@oan.es}}

\pubyear{2015}
\volume{315S}  
\jname{From  interstellar clouds to star-forming galaxies: universal processes?}
\editors{P. Jablonka, P. Andr\'e \& F. van der Tak, eds.}
\begin{document}

\maketitle

\begin{abstract}

Galaxy nuclei are a unique laboratory to study gas flows. High-resolution imaging of the gas flows in galactic nuclei are instrumental in the study of the fueling and the feedback of star formation and nuclear activity in nearby galaxies. Several fueling mechanisms can be now confronted in detail with observations done with state-of-the-art interferometers. Furthermore, the study of gas flows in galactic nuclei can probe the feedback of activity on the interstellar medium of galaxies. Feedback action from star formation and AGN activity is invoked to prevent galaxies from becoming overly massive, but also to explain scaling laws like black hole (BH)-bulge mass correlations and the bimodal color distribution of galaxies. This close relationship between galaxies and their central supermassive BH can be described as {\it co-evolution}. There is mounting observational evidence for the
existence of gas outflows in different populations of starbursts and active galaxies, a manifestation of the feedback of activity. We summarize the main results recently obtained from the observation of galactic inflows and outflows in a variety of active galaxies with current millimeter interferometers like ALMA or the IRAM array. 

\end{abstract}

\firstsection 
\section{Introduction}

Galaxy scaling relations indicate that the growth of supermassive BH and the assembly of galaxy spheroids are closely connected.
In order to understand the physical framework explaining the nature of this  connection we are to study the complex interplay between gas inflows and gas outflows in galaxies.  The gas component can form stars on its way to the BH. Quantifying the efficiency of angular momentum transport in galaxy disks help address how star formation (SF) and nuclear (AGN) activity can be fed and on what timescales this can occur. Furthermore the gas component can also experience significant feedback from SF and AGN activity on its way to the BH. The study of gas outflows, considered to be a smoking gun evidence of ongoing mechanical feedback, is crucial to assess the efficiency of the feedback loop.

Finding a  causal link between the presence of gas inflows/outflows and the existence of ongoing star formation or nuclear activity in galaxies is challenging from the observational point of view because both the timescales and the spatial scales associated with these phenomena can be very different. In particular, although the timescale associated for the total BH-growth phase can be comparable to the SF timescale (10$^{7-9}$ years), the BH growth phase is suspected to be a collection of thousands of bursts events, each of them lasting only for about 10$^5$ years. The emerging picture sees AGN accretion as a chaotic and largely time-dependent process. AGN can {\it flicker} and may oscillate between high-accretion modes, where feedback is dominated by radiation and winds, and low accretion modes, where feedback is 
mostly kinetic and driven by jets (e.g., \cite[Schawinski et al.~2015]{Sch15}; \cite[King \& Nixon~2015]{Kin15}).   

In this paper we discuss how molecular line observations can be used to study the fueling and the feedback of star formation nuclear activity. In particular we will illustrate the importance of high-resolution observations in the millimeter range done with interferometers to study inflows and outflows in nearby galaxies, where the relevant processes responsible for secular evolution can be studied in great detail. 

\section{Tracing molecular inflows}

 The search of a mechanism responsible of feeding nuclear activity is arduous in  low-luminosity AGN (LLAGN). Several 
mechanisms that show unequal success to induce gas inflow have been discussed in the literature. Gravitational barriers can sometimes stop the gas on its way to the nucleus. The case of LLAGN is special as the total BH growth phase in this family of objects is expected to consist of short-lived episodes, each of them likely lasting for less than the lifetime of the agent responsible for feeding the nucleus. Different timescales of potentially more than one mechanism can mask the correlation further (e.g., \cite[Hopkins et al.~2011]{Hop11}). While various modelsÕ predictions are still debated in the literature, there is still ample room for improvement in the picture drawn from observations. Observations that probe the critical scales for AGN feeding ($<10-100$~pc) can help to make a significant progress in the field.

\begin{figure*}[tb!]
\begin{center}
 \includegraphics[width=7.5cm, angle=-90]{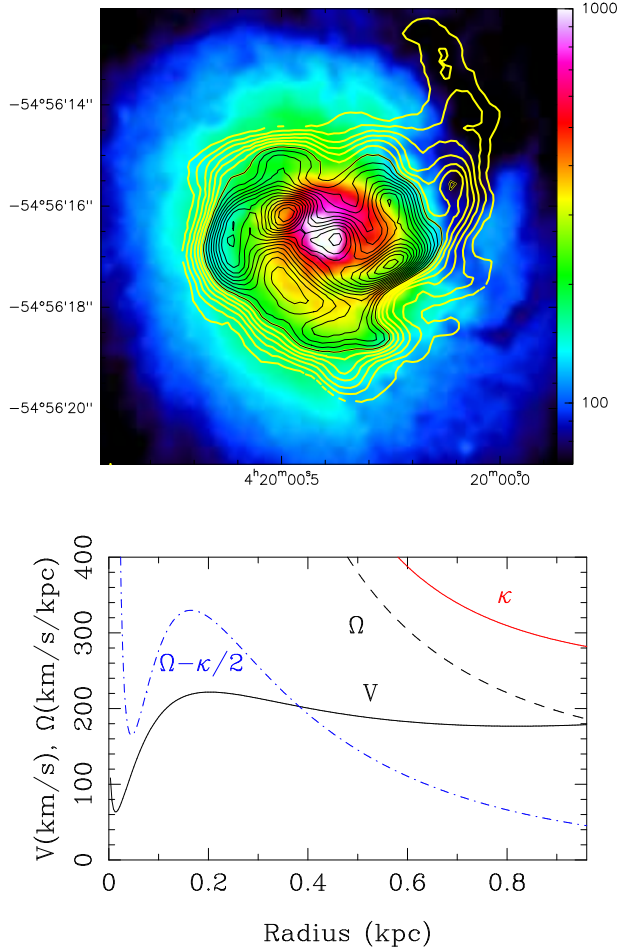}

   \includegraphics[width=5.5cm, angle=-90]{garcia-burillo-fig1b.ps}     
 \caption{{\it (Top panel)}~The CO(3--2) emission contours obtained by ALMA towards the circumnuclear disk of the Seyfert~1 NGC~1566 are overlaid on the (F606W) HST image of the galaxy. The CO image shows a trailing spiral structure from $r\leq50$~pc to $r=300$~pc, which closely follows the dusty spiral seen by the HST.     {\it (Lower panel)}~The rotation curve (black), epicyclic frequency $\kappa$ (red), and $\Omega-\kappa/2$ curve (blue) in NGC~1566. The influence of the  super-massive black hole explains reversal of  $\Omega-\kappa/2$ curve at  $r\leq50$~pc  and implies negative torques on the gas spiraling to the nucleus on these spatial scales. Figures adapted from 
 \cite[Combes et al.~(2014)]{Com14}. } 
   \label{fig1}
\end{center}
\end{figure*}

The Nuclei of Galaxies (NUGA) survey (see \cite[Garc\'{\i}a-Burillo \& Combes~2012]{Gar12} and references therein) is an interferometer CO survey of 25 low-L nearby AGNs conducted with the IRAM array. With these observations we were able to probe the distribution and kinematics of molecular  gas in the central kpc-scale disks in these galaxies with high spatial resolution ($\sim10-100$~pc) to search for evidences of ongoing AGN feeding. Together with the CO NUGA maps, we have access to high-resolution NIR maps obtained by HST, Spitzer and/or ground-based telescopes for this sample. The NIR imaging is used to derive the stellar potentials, which are combined with the CO maps, to derive the gravity torque budget. Two-dimensional torque maps are used  to estimate  the torques averaged over the azimuth $t(r)$ using the gas column density $N(x,y)$ derived from CO as weighting function (\cite[Garc\'{\i}a-Burillo et al.~2005]{Gar05}; \cite[Haan et al.~2009]{Haa09}). With this information at hand we quantitatively evaluated if there were signs of ongoing AGN fueling in the disks of the galaxies analyzed, down to the spatial resolution of our observations.  

About one third of the LLAGNs analyzed in NUGA show negative torques $t(r)$$<$0, indicative of inflow, down to typical radial distances $r$$<$25--100~pc. Among these galaxies, which can be considered as those showing smoking gun evidence of ongoing fueling,  we distinguish three types of  objects: 

{\it --Galaxies showing dynamical decoupling of embedded structures}: these include nuclear stellar bars - within - bars/ovals (e.g.; NGC~2782) and nuclear ovals - within -bars  (e.g.; NGC~4579).

{\it --Galaxies with ILR-free large-scale stellar bars}: these galaxies are characterized by the apparent absence of gravity torque barriers in their nuclei (e.g.; NGC~3627).

{\it --Galaxies with nuclear gas spirals}:  long predicted by theoretical models and seen in dusty structures in HST images, these gas instabilities are started to by seen by ALMA in CO line emission, like in the Seyfert~1 galaxy NGC~1566 (Fig.~\ref{fig1}). The trailing spiral
seen in the central  $r\leq50-300$~pc of NGC~1566  reflects the influence of the BH on the gas response, which facilitates gas inflow down to $r\leq50$~pc.

 On the contrary, about two thirds of the LLAGNs analyzed in NUGA show positive torques $t(r)$$>$0, indicative of outflow,  down to typical radial distances $r$$<$300~pc. This {\it puzzling} gravity torque budget is found in two categories of objects in our sample:

{\it --Galaxies showing Ônon-cooperativeÕ embedded structures}: in these galaxies, nuclear bars  or ovals - within - bars, like those found in NGC~4321 and NGC~6951 are not
always conducive to gas inflow at present. 

{\it --Galaxies showing featureless/mostly axisymmetric potentials}: the absence of a clear non-axisymmetric feature in  the stellar potential makes the angular momentum transfer in the nucleus a very inefficient process in some galaxies at present (e.g.; NGC~4826; NGC~7217; NGC~5953).

The overall results of NUGA indicate that high spatial resolution is instrumental in quantifying angular momentum transfer processes at critical scales ($\sim10-100$~pc). Gas flows in NUGA maps reveal a wide range of large-scale and embedded $m=2, m=1$ instabilities in the circumnuclear disks of LLAGNs. The derived gravity torque maps indicate that gas is frequently stalled in rings, which are the signposts of gravity torque barriers. Only $\sim1/3$ of galaxies show negative torques down to $\sim50$~pc. These various fueling mechanisms are related to bar cycles. 

However, the $r\sim1-10$~pc-scales are to a large extent uncharted territory for molecular line observers. We need more observational constraints to investigate what are the main fueling mechanisms at work on these scales and confirm the predictions of numerical simulations (e.g., \cite[Hopkins 
\& Quataert~2011]{Hop11};  \cite[Hopkins et al.~2015]{Hop15};  \cite[Wada~2001]{Wad01};  \cite[Wada et al.~2009]{Wad09}).

\begin{figure*}[th!]
\begin{center}
 \includegraphics[angle=90, width=6.7cm, height=6.45cm]{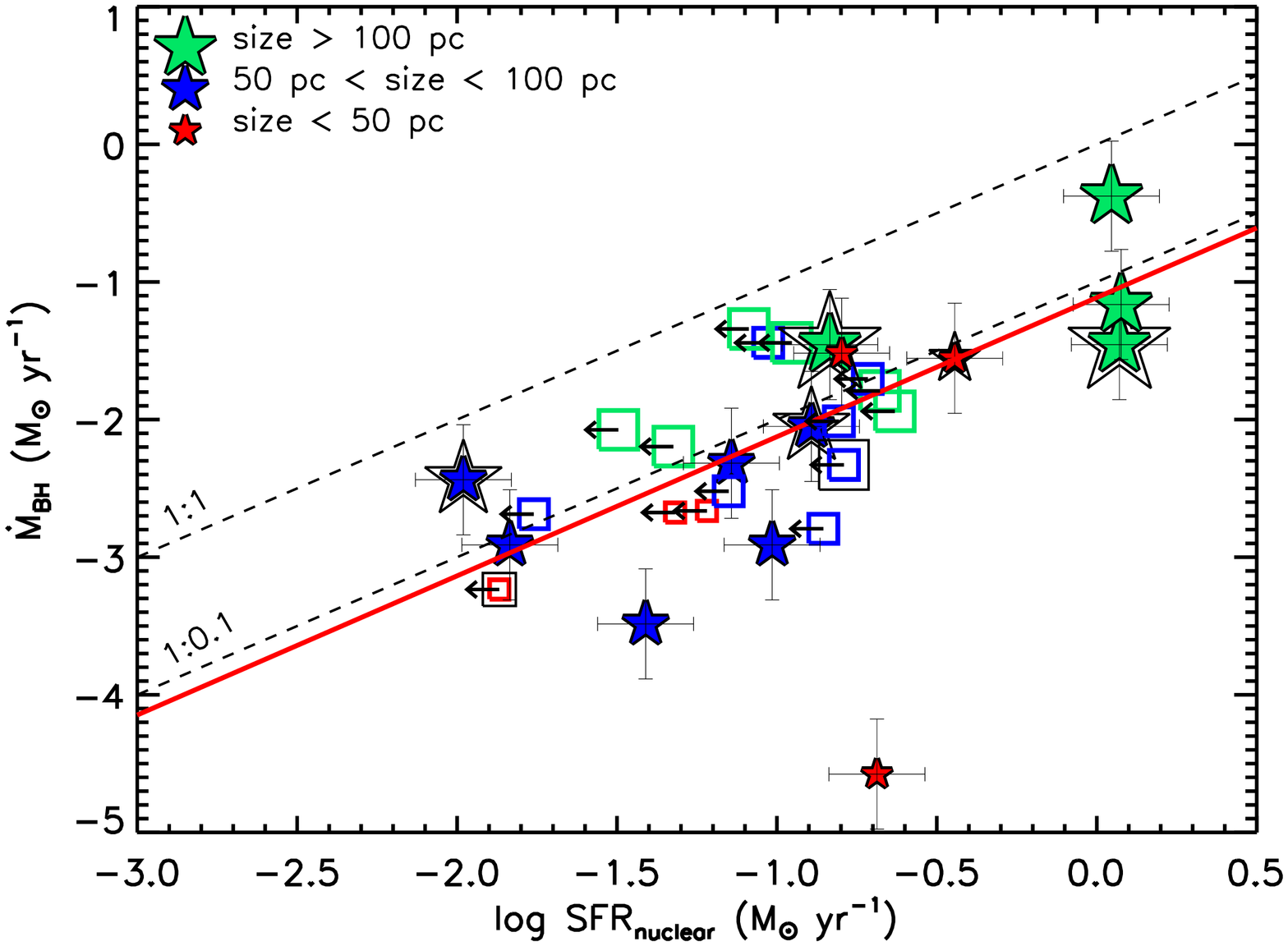}
  \includegraphics[width=6.7cm, angle=0, height=6.7cm]{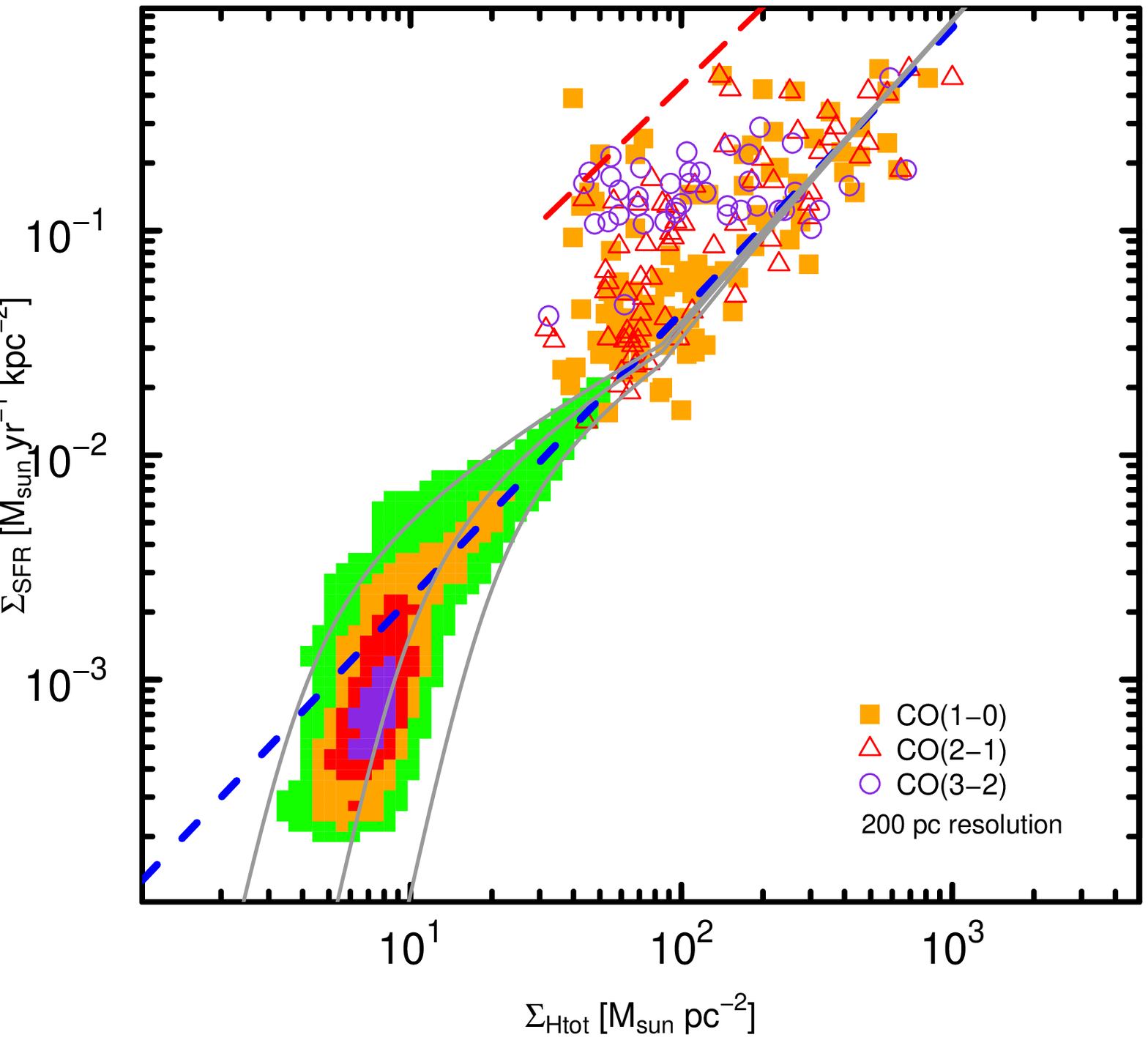} 
 \caption{{\it a)Left panel:}~The nuclear ($r\leq100$~pc) $SFR-dM_{\rm BH}/dt$ scaling relation obtained in the sample of 29 Seyfert galaxies analyzed by 
  \cite[Esquej et al.~(2014)]{Esq14}. The solid line represents the fit to the detections of the nuclear 11.3~$\mu$m PAH feature. The sizes/colors of the symbols indicate the different physical sizes probed.  {\it b)Right panel:}~ The spatially resolved KS-law derived for the four active NUGA galaxies studied by \cite[Casasola et al.~(2015)]{Cas15} at a spatial resolution of 200~pc, compared to the results from \cite[Bigiel et al.~(2008)]{Big08}, and the model predictions from \cite[Krumholz et al.~(2009)]{Kru09}. The blue line identify the branch of the SF galaxies, while the red line identifies the merger sequence of  \cite[Genzel et al.~(2010)]{Gen10}.}
   \label{fig2}
\end{center}
\end{figure*}

\section{The SF-AGN connection}

Many works have tried to establish a causal connection between BH and galaxy growth, by studying to what extent the rate of SF is correlated with the BH feeding rate and, most importantly, by finding on what spatial scales ({\it global}$\sim$kpc or {\it nuclear}$\sim$100~pc) this correlation may hold. Despite the fact that both SF and BH feeding rely on a common cold gas supply, the fact that the timescales associated to SF and AGN feeding can be very different adds non-trivial complications.
\cite[Delvecchio et al.~(2015)]{Del15}  have recently studied the relation between AGN accretion and the global SF rate in a sample of 8600 SF galaxies up to $z=2.5$ from the GOODS and COSMOS surveys, finding a good correlation between these two  quantities, in spite of the large variance due to the different spatial/time-scales.  However the prediction of the recent numerical simulations of \cite[Volonteri et al.~(2015ab)]{Vol15a, Vol15b} are in tension with these results, as they  anticipate that only on {\it nuclear} scales the SF-BH growth rates should show any significant correlation. 

Despite first claims that the $SFR$-$BHR$ correlation may be 'blurred' even on nuclear scales, due to an eventual delay in fueling the AGN after a circumnuclear SF episode in Seyfert galaxies,  (e.g.; \cite[Davies et al.~2007]{Dav07}), \cite[Esquej et al.~(2014)]{Esq14} have recently shown that the  $SFR$-$dM_{\rm BH}/dt$ scaling law obtained using a sample of 29 Seyferts  shows evidence of a significant correlation (Fig.~\ref{fig2}{\it a}), a result that confirms the predictions of numerical simulations. Taken at face value, the results of   \cite[Esquej et al.~(2014)]{Esq14} indicate that SF and BH growth may be running in parallel in a symbiotic way in the circumnuclear regions of LLAGN.
Furthermore, this picture seems to be confirmed by the work of \cite[Casasola et al.~(2015)]{Cas15}, who found that the SF efficiency of molecular gas in the circumnuclear disks of four NUGA targets, analyzed with $\sim200$~pc spatial resolution, lies between the values characterising the sequence of SF galaxies and that of mergers (Fig.~\ref{fig2}{\it b}). However, the question of whether AGN activity can significantly quench SF in AGN-hosts is far from being settled. Using larger samples two different groups (\cite[Shimizu et al.~2015]{Shi15}; \cite[Mullaney et al.~2015]{Mul15}) have found evidence that a significant percentage of AGN hosts 
seem to be located below the main sequence of SF galaxies, an indication that the brightest AGN may be in the process of quenching.

\begin{figure*}[t!]
\begin{center}
 \includegraphics[angle=0, width=8.5cm]{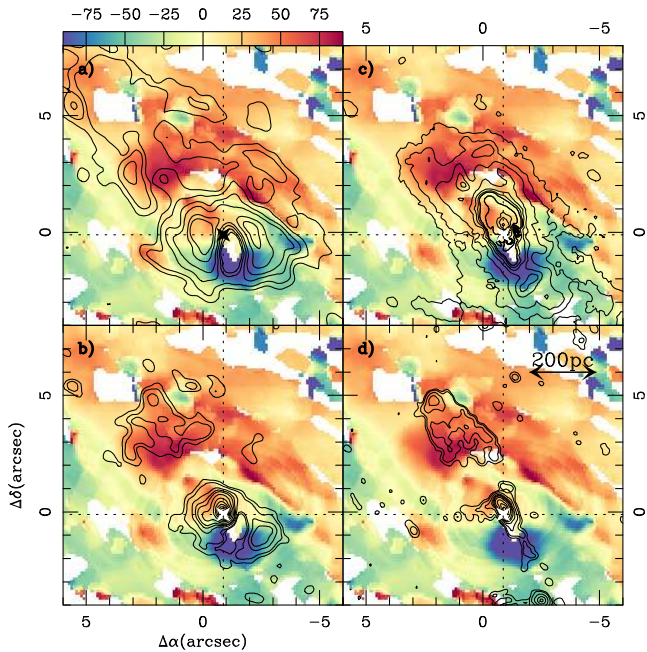}
 \caption{Overlay of the residual velocity field of molecular gas in NGC~1068 observed by ALMA (in color scale) obtained after subtraction of the model rotation curve, with the contours representing: the integrated intensity of CO(3--2) ({\it a)upper left panel}),  the 349~GHz continuum emission  ({\it b) lower left panel}), the HST Pa$\alpha$ emission ({\it c) upper right panel}), and the 22~GHz VLA map of \cite[Gallimore et al.~(1996)]{Gal96} ({\it d) lower right panel}). Figure from \cite[Garc\'{\i}a-Burillo et al.~(2014)]{Gar14}.}
   \label{fig3}
\end{center}
\end{figure*}

\section{Tracing molecular outflows}

Outflows can prevent galaxies from becoming exceedingly massive and help regulate the fueling of activity. There is observational evidence for the existence of gas outflows in different populations of nearby galaxies, including ultra luminous infrared galaxies (ULIRGs), radio galaxies, quasars and Seyferts. The outflow phenomenon concerns all the phases of the ISM.  In particular, the study of the molecular component, being the most massive phase of the neutral ISM in the central kiloparsec regions, is crucial if we are to quantify the impact of outflows on galaxy evolution.

The quest for molecular outflows using mm-interferometers like ALMA or the IRAM array has yielded an increasing number of detections in ULIRGs (e.g., \cite[Feruglio et al.~2010, 2013, 2015]{Fer10, Fer13, Fer15};  \cite[Aalto et al.~2012, 2015]{Aal12, Aal15}; \cite[Cicone et al.~2012, 2014]{Cic12, Cic14};  \cite[Garc\'{\i}a-Burillo et al.~2015]{Gar15}), radio galaxies (e.g., \cite[Tremblay~2014]{Tre14}) and LLAGN
 (e.g., \cite[Alatalo et al.~2011,2015]{Ala11, Ala15}; \cite[Combes et al.~2013]{Com13}; \cite[Morganti et al.~2013, 2015]{Mor13, Mor15}; \cite[Garc\'{\i}a-Burillo et al.~2014]{Gar14}; \cite[Sakamoto et al.~2014]{Sak14}).
The high-resolution capabilites of these type observations are instrumental in constraining the geometry as well as the mass, momentum and energy budgets of the molecular outflows ($dM_{\rm out}/dt$, $dP_{\rm out}/dt$, $L^{\rm kin}_{\rm out}$). Having good estimates of these three quantities  is a prerequisite in order to study the specific role that star formation versus AGN feedback have in launching and maintaining molecular outflows.

\begin{figure*}[tb!]
\begin{center}
 \includegraphics[angle=0, width=14.0cm]{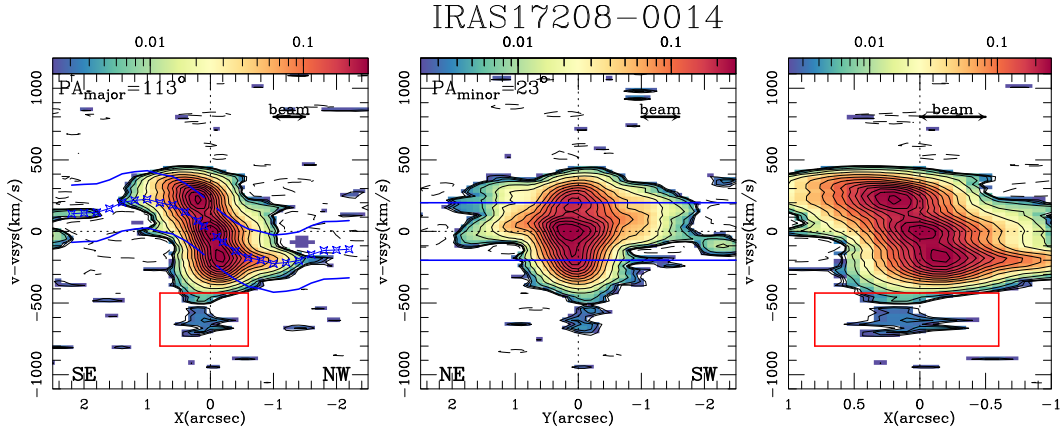}
 \caption{{\it a)Left panel:}~The CO(2--1) p-v plot taken along the kinematic major axis of IRAS~17208-0014 derived with the IRAM array. We delineate  the (projected) rotation curve and the allowed virial range around it.  Similarly, we delimit the region where emission is detected from the {\it line wing} betraying the most extreme blue-shifted velocities of the outflow  ($v-v_{\rm sys}$~=~[--450,--750]~km~s$^{-1}$).
 {\it b)Middle panel:}~The p-v plot taken along the kinematic minor axis. {\it c)Right panel:}~Same as  {\it left panel} but zooming in on the central $\pm1"$ region around the center.   Figure taken from \cite[Garc\'{\i}a-Burillo et al.~(2015)]{Gar15}.}
   \label{fig4}
\end{center}
\end{figure*}

\begin{figure*}[tbh!]
\begin{center}
 \includegraphics[angle=0, width=7.0cm]{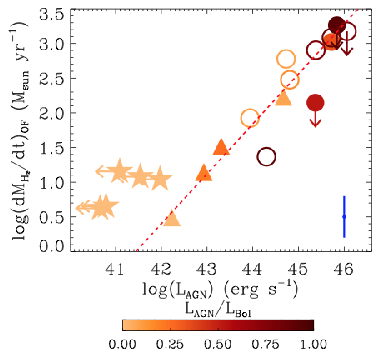}
 \includegraphics[width=6.7cm, angle=0, height=6.5cm]{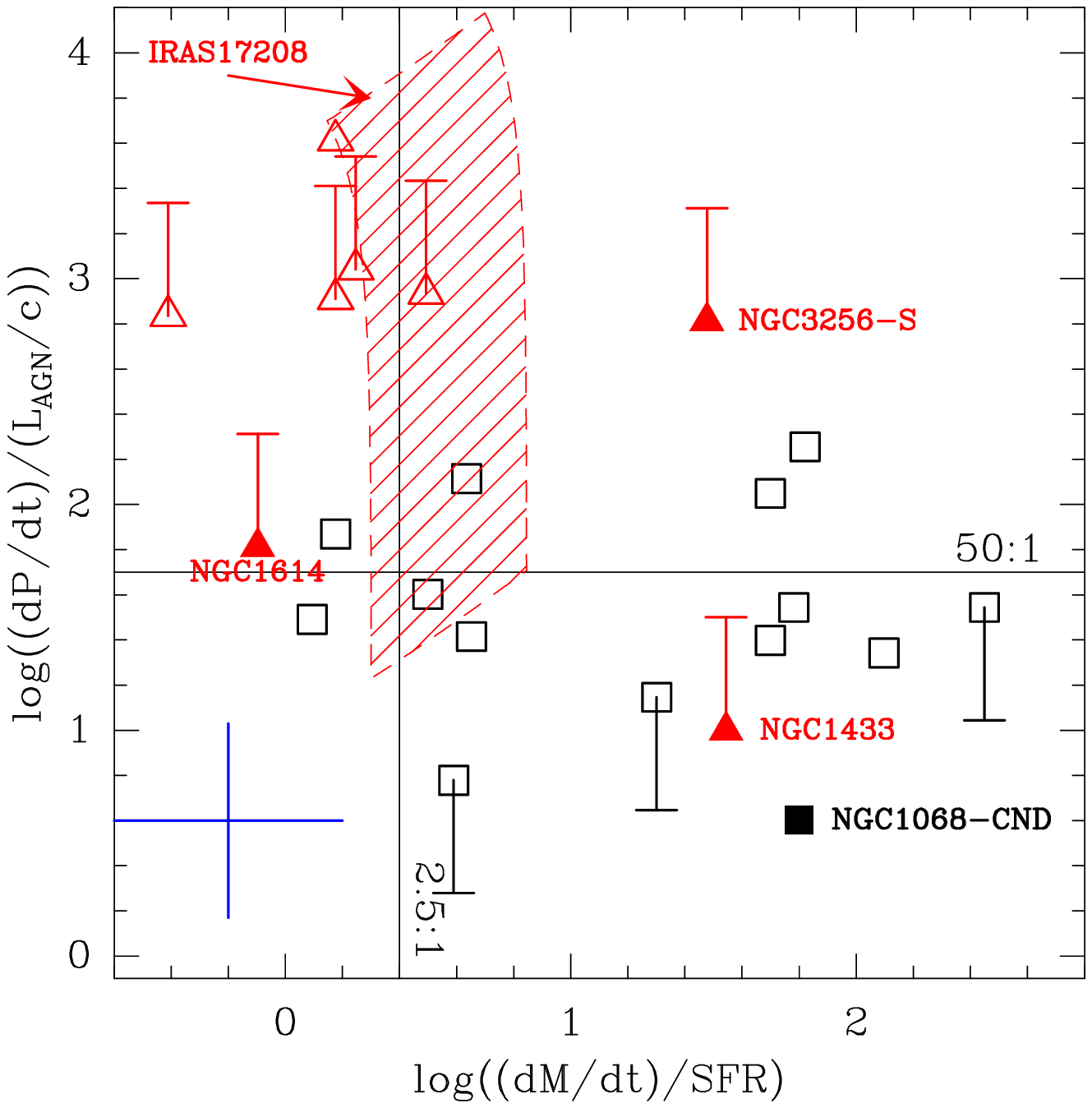} 
 \caption{{\it a)Left panel:}~Outflow mass rate ($dM/dt$) as a function of the AGN bolometric luminosity $L_{\rm AGN}$ for the sample of galaxies analyzed by \cite[Cicone et al.~(2014)]{Cic14}, who used CO observations to trace molecular gas masses. Filled and open circles represent unobscured and obscured AGNs respectively, LINERs are plotted as triangles and ÒpureÓ starbursts as stars.  The red line illustrates the fit to the deteted AGNs. Figure taken from \cite[Cicone et al.~(2014)]{Cic14}. {\it b)Right panel:}~Momentum boost factor $(dP/dt)/(L_{\rm AGN}/c)$ as a function of mass-loading factor ($(dM/dt)/SFR$) for the galaxies compiled by \cite[Garc\'{\i}a-Burillo et al.~(2015)]{Gar15}. Open (red) triangles show the data from pure starburst galaxies; open (black) squares identify the data of low and high-luminosity AGNs. Filled symbols identify the data from the new additions. The location of galaxies in this diagram help identify molecular outflows driven by SF (upper left corner), radiation pressure or winds of AGN (lower right corner),  AGN jets (upper right corner) or a combination thereof in mixed/undefined cases (lower left corner). Figure adapted from \cite[Garc\'{\i}a-Burillo et al.~(2015)]{Gar15}.}
   \label{fig5}
\end{center}
\end{figure*}

As an illustration of the power of mm-interferometry, the sensitivity and spatial resolution of ALMA have provided  an unprecedented view of the distribution and kinematics of the dense molecular gas  in the central $r\sim2$~kpc of the Seyfert 2 galaxy NGC~1068 with spatial resolutions $\sim0.3"-0.5"$ ($\sim20-35$~pc) (\cite[Garc\'{\i}a-Burillo et al.~2014]{Gar14}). The gas kinematics from $r\sim50$~pc out to $r\sim400$~pc reveal a massive outflow in all molecular tracers mapped by ALMA. The tight correlation between the ionized gas outflow, the radio jet, and the occurrence of outward motions in the disk suggests that the outflow is AGN driven (Fig~\ref{fig3}).  The outflow rate estimated in the circumnuclear disk (CND), $\mathrm{d}M/\mathrm{d}t\sim63^{+21}_{-37}~M_{\odot}$~yr$^{-1}$, is an order of magnitude higher than the star formation rate at these radii. The molecular outflow could quench star formation in the inner $r\sim400$~pc of the galaxy on short timescales of $\leq1$~Myr and regulate gas accretion in the CND.

Galaxy evolution scenarios foresee that the feedback of SF and AGN activity can drive the transformation of gas-rich spiral mergers into ULIRGs, and, eventually, lead to the build-up of QSO/elliptical hosts. In an attempt to address the role that star formation and AGN feedback have in launching and maintaining the molecular outflows in starburst-dominated advanced mergers, \cite[Garc\'{\i}a-Burillo et al.~(2015)]{Gar15} used the IRAM array to image with high spatial resolution ($0.5"-1.2"$) the CO(1--0) and CO(2--1) line emissions in NGC~1614 and IRAS~17208-0014, respectively.  These observations have detected in both mergers the emission from high-velocity {\it line wings} that extend up to $\pm$500--700~km~s$^{-1}$, well beyond the estimated virial range associated with rotation and turbulence, a signature of outflows (Fig.~\ref{fig4}).  In IRAS~17208-0014, gas emission from the blueshifted {\it line wing}  is kinematically decoupled from the disk: its apparent  kinematic major axis is reversed. This pattern can be explained by a non-coplanar molecular outflow. This scenario is reinforced after subtraction of the rotation curve, which reveals that the kinematic decoupling of the outflow starts at $\mid v-v_{\rm sys}\mid$~$\geq300$~km~s$^{-1}$. The mass, energy and momentum budget requirements of the molecular outflow in IRAS~17208-0014 can be accounted for by the existence of a so far undetected ({\it hidden}) AGN of $L_{\rm AGN}\sim7\times10^{11}~$L$_{\odot}$. 

Several diagnostic tools are used in the literature to study the location of  different populations of galaxies where molecular outflows have been detected (\cite[Cicone et al.~2014]{Cic14};  \cite[Garc\'{\i}a-Burillo et al.~2015]{Gar15}). These type of diagrams are crucial if we are to identify the driving agent of the outflows: SF versus AGN, but also, more specifically, radiation pressure, winds or jets (Fig~\ref{fig5}).
 Based on their compiled sample of 19 sources,  \cite[Cicone et al.~(2014)]{Cic14} found that star formation can drive outflows with d$M$/d$t$ values up to 2--4 times the value of the star formation rate ($SFR$). However, the presence of an AGN, even if weak ($L_{\rm AGN}/L_{\rm bol} < $~0.10),  can strongly boost an outflow with   d$M$/d$t \propto L_{\rm AGN}$. In the most extreme quasars ($L_{\rm AGN}/L_{\rm bol} \geq 0.80$), outflow rates can be up to $100 \times SFR$, and therefore have a decisive  impact on the fueling of the star formation and nuclear activity of their hosts. This picture is similar to the scenario derived from the numerous observations of outflows detected in the OH and NaID lines (\cite[Fischer et al.~2010]{Fis10}; \cite[Sturm et al.~2011]{Stu11}; \cite[Rupke \& Veilleux~2013]{Rup13}; \cite[Spoon et al.~2013]{Spo13}).
 
There is still a number of key results emerging from observations that remain to be understood.  Molecular outflows are energetically relevant, especially, but not exclusively, in luminous starbursts and AGN. In the case of Seyferts,  molecular outflows, although less extreme in their properties, play a relevant role in regulating the fueling of activity.  As a general feature, the velocities measured in molecular outflows suggest that most of the gas is likely to fountain back to the disk. Furthermore, the radio-mode feedback seems to be at work not only in extreme radio-loud systems, but also in Seyfert galaxies. This may be connected to the changing accretion/feedback modes in {\it flickering} AGN.



\end{document}